\documentclass[twocolumn,showpacs,preprintnumbers,amsmath,amssymb,prc]{revtex4-1}
\usepackage{epsfig}
\newcommand{\bd}[1]{ \mbox{\boldmath $#1$}  }

\begin{document}

\title{Three-body properties of low-lying $^{12}$Be resonances}

\author{E. Garrido$\:^1$, A.S. Jensen$\:^2$, D. V. Fedorov$\:^2$, J. G. Johansen$\:^2$}
\affiliation{$^1$ Instituto de Estructura de la Materia, IEM-CSIC,
Serrano 123, E-28006 Madrid, Spain}
\affiliation{$^2$ Department of Physics and Astronomy, Aarhus University, 
DK-8000 Aarhus C, Denmark} 

\date{\today}

\begin{abstract}
We compute the three-body structure of the lowest resonances of
$^{12}$Be considered as two neutrons around an inert $^{10}$Be core.
This is an extension of the bound state calculations of $^{12}$Be into the 
continuum spectrum.  We investigate the lowest resonances of angular
momenta and parities, $0^{\pm}$, $1^{-}$ and $2^{+}$.  Surprisingly
enough, they all are naturally occurring in the three-body model.  We
calculate bulk structure dominated by small distance properties as
well as decays determined by the asymptotic large-distance
structure.  Both $0^{+}$ and $2^{+}$ have two-body $^{10}$Be-neutron $d$-wave
structure, while $1^{-}$ has an even mixture of $p$ and
$d$-waves. The corresponding relative neutron-neutron partial waves
are distributed among $s$, $p$, and $d$-waves. The branching ratios
show different mixtures of one-neutron emission, three-body direct, and
sequential decays. We argue for spin and parities, $0^{+}$, $1^{-}$ and
$2^{+}$, to the resonances at $0.89$, $2.03$, $5.13$, respectively. The
computed structures are in agreement with existing reaction measurements.
\end{abstract}


\maketitle

\section{Introduction}
Cluster and halo states in light nuclei have been studied for several
decades \cite{tan85,han87,zhu93,jen04}. These structures are to a large extent decoupled from
more complicated many-body states since they occupy essentially
separate parts of the Hilbert space.  The descriptions of these cluster
structures are usually in terms of few-body correlations, i.e., two
or three almost inert clusters interacting through effective potentials.
These constituent clusters are themselves nuclear systems.  Thus, the
full nuclear many-body system is described by separate relative and
internal cluster degrees of freedom.  The first and most important
step for halo states is then to freeze the internal motion in a given
structure, and solve the remaining relative motion.

Bound states are only a small part of the possible quantum structures
of a given many-body system. Among the infinitely many continuum
states one can find resonances, still by definition in the continuum, but that
can be understood as a
discrete continuation of the set of bound state structures.  As the
excitation energy increases it is expected that the underlying
assumption of uncoupled internal and relative cluster motion becomes
increasingly violated.  However, a number of low-lying many-body
resonances in light nuclei can still be well described as cluster structures
even when they appear in the continuum.  Prominent examples are found
in stable nuclei like $^{12}$C and $^{9}$Be \cite{kan98,des02,nef04,alv08,
gor95,sum02,bar06,alv08b,gar10}, but even ground
states like in $^{6}$Be \cite{cso94,vas01,gar07,pap10} and $^{10}$Li \cite{zin95,gar02,jep06}  
may appear as resonances.

A number of beta unstable light nuclei exhibit particle stable bound
cluster states, f.ex. $^{11}$Li \cite{zhu93,gar95}, $^{17}$Ne \cite{gim98,gar04,ois10} 
and $^{12}$Be \cite{ots93,nun96,rom07b,duf10,kan10}.  In many cases the bound 
states are well established, but very often their particle unstable spectrum 
is much less known due to practical experimental difficulties. The excited
$1^-$ state in $^{11}$Li is a clear example \cite{kor97,kum02,gar02b,nak06}.

An interesting point is that very often the neutron dripline has states of 
a unique structure. In particular,
$^{11}$Be is a nucleus with two bound states, both with a two-body halo
structure. The ground state of $^{12}$Be is rather well bound but
three halo states appear as excited states, and $^{13}$Be is particle
unbound, whereas $^{14}$Be has one particle-bound state \cite{des95,sim07}. The
complicated dripline structure for the Be-isotopes is primarily due to
the second $s$-state intruder and eventual inversion with the
$p$-state in $^{11}$Be \cite{tal60}.  Correlations of the valence neutrons in a
larger space result in unusually many bound states compared to other
light dripline nuclei.

Beside the bound (halo) states in $^{12}$Be, a $0^{-}$-state has been
suggested \cite{rom07a,blan10}, looked for in experiments, but so far not 
found as a bound state.
The structure of the effective potentials in the three-body problem of
$^{12}$Be ($^{10}$Be$+n+n$) is complicated and suggesting more excited
states than the known bound states.  The one-neutron threshold is
lower than the two-neutron threshold, implying that a resonance of
one-body nature could appear between the two thresholds. Its
structure would then at large distance have to be one neutron far away
from $^{11}$Be either in the ground or the excited state.

There has been very little experimental investigations of the
resonances in $^{12}$Be. Two resonances were seen in a $^{10}$Be(t,p)$^{12}$Be
experiment by Fortune et al. \cite{for94}, which was later confirmed by Bohlen et al. \cite{boh08}. 
The energy and width were measured, and set to $0.89$~MeV and $2.03$~MeV
above the two-neutron threshold. The strong population of the resonances in a two neutron
transfer indicates a natural parity $(0^+,1^-,2^+,..)$, and tentatively quantum numbers were given from DWBA calculations.

It seems to be very appropriate to extend the theoretical bound state
study of $^{12}$Be to the continuum.  The purpose of the present paper
is to investigate the low-lying resonance structure of $^{12}$Be.  In
section~II we very briefly sketch the method, the previously
applied effective $^{10}$Be-neutron interactions, and the adiabatic 
$^{12}$Be potentials. In section~III, we present
resonance energies and their quantum numbers.  In section~IV, we discuss
the resonance structures. In section~V, we discuss the production and decay modes
of the resonances and compare our results with known experimental data. Finally, in section VI we
give a short summary and the corresponding conclusions.

\section{Basic ingredients}
The theoretical framework has been previously well described and 
successfully applied \cite{rom07b}. We only briefly sketch to define notation
and interactions. The resulting adiabatic potentials are the basic
ingredients in the subsequent calculations.

\subsection{Method and interactions}

We use the adiabatic hyperspherical expansion method to calculate the
three-body properties of the system in question, that is $^{12}$Be
($^{10}$Be$+n+n$).  This method strictly deals with three entities
treated as point-like particles interacting pairwise with each other
\cite{nie01}.  The Faddeev equations are first formulated in
hyperspherical coordinates which consists of the hyperradius, $\rho$,
and five angles collectively named $\Omega$.

Assuming that the two-body interactions are known, we can solve the
angular part of the Faddeev equations for fixed hyperradius. Due to
the restriction of finite intervals for all five angular coordinates
the eigenvalue spectrum is discrete, although in principle with
infinitely many elements.  The angular solutions form a complete set
which is exploited as a basis for expansion of the total
wave function. This provides finally a coupled set of one-dimensional
hyperradial equations.  The details of the method are discussed in
\cite{nie01}.

Bound states are solutions to the radial equation with an
exponentially falling radial large-distance behavior.  Resonances are
solutions to the same hyperradial equations but with boundary
conditions corresponding to only outgoing waves. We apply the complex
rotation method on the hyperspherical coordinates, that is
only the hyperradius is scaled by $\exp(i\theta)$ \cite{fed03}. The resonance
boundary conditions are then transformed to an exponential fall-off,
precisely as the bound states, provided the rotation angle, $\theta$,
exceeds $\arctan(\Gamma/2E_R)$, which is the value corresponding to rotation of
the real-energy axis to the position of the resonance,
$(E_R,E_I=-\Gamma/2)$.  We have here denoted the real and imaginary
values of the resonance energy by $E_R$ and $E_I$. 

The two-body interactions between the two neutrons and between neutron
and $^{10}$Be can be chosen through different criteria.  They are all
related to the corresponding two-body properties of bound states
and/or resonances. The crucial pieces are the $s$, $p$, and $d$-wave
interactions.  The neutron-neutron interaction from \cite{rom07b}
reproduces low-energy scattering properties for the different partial
waves.  For neutron-$^{10}$Be, we choose effective potentials for each
set of quantum numbers such that the lowest computed energy reproduces
the bound state or resonance energy.  

In particular, we use the interaction labeled I in \cite{rom07b}, where the
central and spin-orbit radial potentials are assumed to have a gaussian shape. 
The range of this interaction for all the partial waves is taken equal to 
3.5 fm, that is the sum of 
the rms radius of the core and the radius of the neutron. For $s$-waves 
the strength is fixed to
fit the experimental neutron separation energy of the $1/2^+$-state in
$^{11}$Be ($-0.504$ MeV). For $p$-waves the two free
parameters (central and spin-orbit strengths) are adjusted to
reproduce the experimental neutron separation energy of the
$1/2^-$-state in $^{11}$Be ($-0.184$ MeV), and
simultaneously push up the $3/2^-$ state, which is forbidden by the
Pauli principle, since it is occupied by the four neutrons in the
$^{10}$Be core.
For the $d$-states it is well established that $^{11}$Be has a $5/2^+$
resonance at 1.28 MeV (energy above threshold), and the most likely
candidate as spin-orbit partner of the 5/2$^+$ state is the known
3/2$^+$-resonance at 2.90 MeV (above threshold) \cite{fuk04}.  
Simultaneous matching of these two resonances (and the corresponding widths)
leads to central and spin-orbit radial potentials for $d$-waves made as a sum of two
gaussians. Further details about the interaction are discussed in \cite{rom07b}. 

Typically, in a three-body calculation an effective three-body force is
introduced in the coupled set of radial equations in order to fine tune the properties 
of the three-body system \cite{nie01}. This
potential can be understood as the way to take care of all those effects that go 
beyond the two-body interactions.
In our calculations we have used a gaussian three-body force, whose
range has been taken equal to 4.25 fm, that is the hyperradius
corresponding to a $^{10}$Be-core and two neutrons touching each
other. The strength of the gaussian is used to adjust the energy
of the computed resonance. Once this is done, the width is 
determined by the potential barrier (height and thickness) at that precise energy.
A change of the range in the three-body potential could affect the properties of 
the barrier (and therefore the width of the resonance), but a modest change  produces 
only very modest variations of the width. This is because the strength is adjusted 
to maintain the energy which is the crucially important quantity
determining the width. Thus the conclusions reached in this work would still hold.

\subsection{Adiabatic cluster potentials}

\begin{figure}
\begin{center}
\epsfig{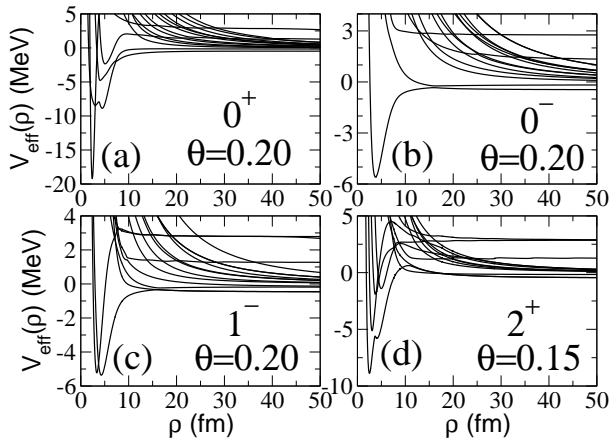}
\end{center}
\caption{Real parts of the dominating lowest-lying adiabatic
  potentials as functions of hyperradius for $^{12}$Be for total
  angular momentum $J^{\pi}= 0^{\pm},1^{-},2^{+}$.  The scaling angle,
  $\theta$ (in rad), is given in the figure for each $J^{\pi}$. }
\label{fig1}
\end{figure}

The solutions to the complex rotated angular equations provide the 
set of complex hyperradial adiabatic potentials whose real parts are shown 
in Fig.\ref{fig1} for each of
the most interesting sets of angular momentum and parity.  Most of
these potentials approach zero at large distances. However, several of
them approach values less than or larger than zero.  These
asymptotic energies are two-body bound state or resonance energies
corresponding to two particles in those relative states while the
third particle is far away. Below zero it must be the $s_{1/2}$ or
$p_{1/2}$ bound states of $^{11}$Be at respectively $-0.504$~MeV and
$-0.184$~MeV.  Above zero it must be the $d_{3/2}$ or
$d_{5/2}$-resonances, respectively at $2.90$~MeV and $1.28$~MeV above
the three-body threshold \cite{fuk04}.

At smaller distances, all the potentials diverge towards $+\infty$,
and at finite, but relatively small distances, a substantial amount of
structure is present. The attractive pockets are responsible for
structures like three-body bound states or resonances. If all
potentials are repulsive at all distances, no structure can arise and
all three particles would try to get as far from each other as
possible.

The $0^{-}$ set of potentials are the simplest as only one of the
potentials exhibits some attraction.  However, this level crosses, or
rather avoids crossing, another purely repulsive level when $\rho$ is
about 15~fm. The large-distance asymptotics of these two crossing levels then
correspond to the $^{11}$Be in the $s_{1/2}$ or $p_{1/2}$ bound states
and the last neutron spatially far away corresponding in
$p_{1/2}$ or $s_{1/2}$-states with respect
to the center of mass of the $^{11}$Be states. Only these combinations
are allowed since the total angular momentum and parity, $0^{-}$, has
to arise after coupling of these angular momenta.  Thus, the
attractive pocket seems to be mostly of $^{11}$Be($p_{1/2}$)
character, since this configuration is reached by the smooth
continuation of the attractive structure to large distances beyond the avoided
crossing.

The $1^{-}$ set of potentials are also relatively simple with only two
potentials with attractive pockets where the thickest and deepest
resembles the lowest $0^{-}$ potential. Also for $1^{-}$, this
potential avoids crossing another potential, and together they form
the same large-distance asymptotic structure as the two lowest $0^{-}$
potentials.  The only difference is that the $s_{1/2}$ and $p_{1/2}$
angular momenta now are coupled to $1^{-}$. The other potential with
an attractive pocket is thinner and rather steeply increasing to
``avoid crossing'' a number of other potentials where the first is the
purely repulsive potential ending up as the lowest at large distance.
This potential therefore has a barrier, and
consequently it may be able to hold a resonance.  Due to the possibly
complicated rearrangements of structures at crossings the decay path
and resulting decay channels cannot be derived by inspection of these
potentials.

Both $0^{+}$ and $2^{+}$ sets of potentials are much more structured.
Now four potentials have attractive pockets at small distances and
each has fast small-scale variation arising from crossings at these
hyperradii. The two large-distance $^{11}$Be structures are again for
both $0^{+}$ and $2^{+}$ found as potentials approaching $-0.504$~MeV
and $-0.184$~MeV. For $0^{+}$ the last neutron is then in $s_{1/2}$
or $p_{1/2}$ states relative to $^{11}$Be($s_{1/2}$) or
$^{11}$Be($p_{1/2}$).  For $2^{+}$, three negative states appear at
large distance with the last neutron in the $d_{5/2}$, $d_{3/2}$ or in
the $p_{3/2}$ state relative to the two bound states of $^{11}$Be.  The
complicated structures of the potentials do not allow quick
predictions of occurrence of bound states or resonances and their
structure or decay properties. Detailed investigations must be carried
out.

\section{Bound state and resonance energies}

The potentials in Fig.\ref{fig1} are applied to the coupled set of
hyperradial equations.  Bound state energies are then first obtained
as described in details in \cite{rom07b}, that is the experimentally
known ground state of $0^{+}$, and the excited states of $0^{+}$, $1^{-}$
and $2^{+}$. In precisely the same framework a $0^{-}$ state was also
found \cite{rom07a}. So far, this state has not been seen in experiments.

The same basic interactions that produced the adiabatic potentials in
Fig.\ref{fig1} are also able to support resonances. They are computed
by the complex scaling technique \cite{fed03} as poles of the $S$-matrix.
The precise positions cannot be reliably obtained since polarization
effects are excluded beyond the two-body level, and contributions from
other neglected degrees of freedom can be crucial. To mock up effects
of these omissions we tune the real parts of the three-body energies
to an a priori chosen value.  The imaginary part, or equivalently the
width of the states, then follow without further adjustment. 

\begin{figure}
\begin{center}
\epsfig{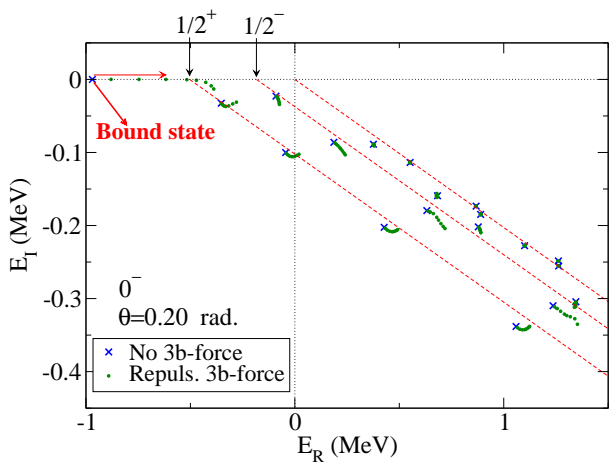}
\end{center}
\caption{(Color online) The complex scaled spectrum for 0$^-$ states in $^{12}$Be
  ($^{10}$Be+n+n).  The down sloping lines are the rotated one-neutron
  thresholds for ground and excited states of $^{11}$Be-neutron and
  the two-neutron threshold for $^{10}$Be+n+n.  The rotation angle is
  0.20 rad.  The points are the discrete states computed in the
  continuum.  The crosses have been obtained without inclusion of any 
three-body force. The close-lying sets are sequences of points arising from
  variation of the strength of the repulsive three-body interaction.  }
\label{fig2}
\end{figure}

The attractive pocket for $0^{-}$ in Fig.\ref{fig1} is sufficiently
strong to support the bound state suggested in \cite{rom07a}, although
no experimental evidence of it has been found so far. The reason for this
can be either because its population in reactions is extremely small, or
it is hidden behind the other states, or it is for some reason pushed
up into the continuum.  We investigate consequences of the last option
where both one-body ($^{11}$Be + n) and two-body ($^{10}$Be
+ n +n) continuum  structures in principle are possible. 

In Fig.\ref{fig2} we show the $0^{-}$ complex energy spectrum where
the repulsion of the three-body interaction is increased from zero.
The spectrum with zero repulsion is shown by the crosses in the 
figure, where a $0^{-}$ bound state appears at an energy of about $-1$ MeV. 
The closed circles show the spectrum when the three-body repulsion
is gradually increased. We see that the first threshold, ground
state of $^{11}$Be, is approached when the real part of the energy
corresponds to a true bound state and the imaginary part is zero.
Passing the threshold allows a finite imaginary part of the energy,
which very quickly increases, and very soon disappears in the rotated
continuum threshold. The result is that this emerging neutron-$^{11}$Be one-body
resonance state is dissolved in this continuum.  It can then stay in
the continuum with a large width or continue to rotate and
end up as a virtual state on another Riemann sheet with zero imaginary
energy. We cannot decide with the present amount of information.  In
any case, we cannot move the resonance to even higher values,
approaching or passing, the next two thresholds.  Thus, we conclude
that the $0^{-}$ state is either bound, a very broad resonance
structure, or a virtual state only revealing itself via an attractive
potential.

\begin{figure}
\begin{center}
\epsfig{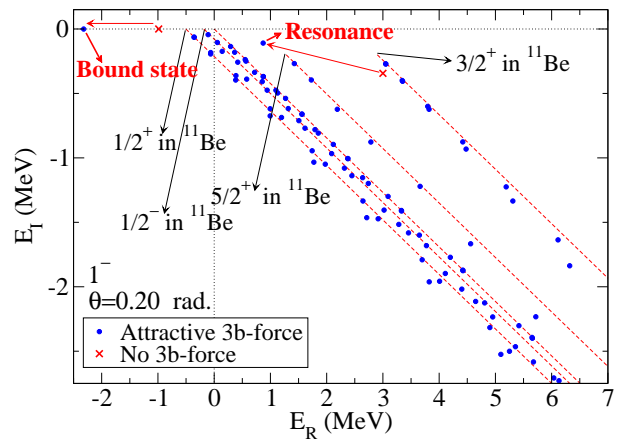}
\end{center}
\caption{(Color online) The complex scaled spectrum for the 1$^-$ states in $^{12}$Be
  ($^{10}$Be+n+n).  The two first down sloping lines are the rotated
  one-neutron thresholds for ground and excited states of
  $^{11}$Be-neutron and the third tilted line is the two-neutron
  threshold for $^{10}$Be+n+n.  The last two tilted lines are the
  thresholds corresponding to the two $d$-wave resonances in $^{11}$Be
  The rotation angle is 0.20 rad.  The points are the discrete states
  computed in the continuum. The crosses are the bound state and the resonance
  obtained without inclusion of any three-body force. The closed circles have been 
  obtained with an attractive three-body force which moves the bound state
  and the resonance as indicated by arrows. }
\label{fig3}
\end{figure}

In Fig.\ref{fig3} we show the $1^{-}$ complex energy spectrum in $^{12}$Be. Now,
together with the three down sloping lines for the one-neutron and two-neutron threshold, 
we also show the two lines corresponding to the rotated thresholds
for the $d_{5/2}$ and $d_{3/2}$ resonances in $^{11}$Be. These two thresholds are
present in the $1^-$ states, which makes the spectrum more complicated than
for the $0^{-}$ states. When no three-body interaction is used, the $1^-$ spectrum
gives rise to one bound state close to $-1$ MeV, and to a resonance at a complex energy 
of $(2.99,-0.34)$ MeV, which are indicated by the crosses in the figure.
The signature of the resonance is a numerically stable and distinguishable point outside all 
the rotated continuum thresholds. An additional attraction, in particular the one provided
by an effective three-body force, moves down the resonance towards the three-body threshold.
It is then not difficult to adjust this attractive force to fit the known experimental
resonance energies in $^{12}$Be. When adjusting the energy to 2.0 MeV the computed width is of about
0.50 MeV, clearly larger than the corresponding experimental width of 0.09 MeV given in \cite{for94}.
When fitting the energy to 0.9 MeV, the computed width is of 0.22 MeV, a factor of 2 larger than the
experimental width quoted in \cite{for94} for this resonance. This last spectrum is the one
shown by the closed circles in Fig.~\ref{fig3}. This size order is in general not allowing
for contributions from configurations omitted in the three-body model.

\begin{figure}
\begin{center}
\epsfig{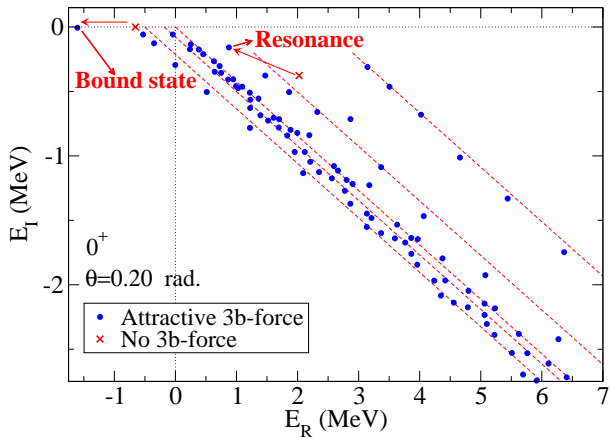}
\end{center}
\caption{(Color online) the same as Fig.\ref{fig3} for the 0$^+$ spectrum
   in $^{12}$Be ($^{10}$Be+n+n). The down sloping lines are the same thresholds as
  described in Fig.\ref{fig3}.  The rotation angle is 0.20 rad.  The
  points are the discrete states computed in the continuum. Again, 
  the crosses are the bound state and the resonance
  obtained without inclusion of any three-body force, and the closed circles have been
  obtained with an attractive three-body force which moves the bound state
  and the resonance as indicated by arrows.}
\label{fig4}
\end{figure}

In Fig.\ref{fig4} we show the $0^{+}$ complex energy spectrum which
gives rise to two bound states (only one shown in the figure).  However, both 
the potentials in
Fig.\ref{fig1} and the present spectrum are more complicated than
those of the $0^{-}$ states.  The properties of the spectrum are
qualitatively very similar to those described for the $1^{-}$ states,
but now the computed resonance energy without three-body potential is 
1.97~MeV. The crosses in the figure indicate this resonance and one of the
bound states. This resonance energy already matches one of the experimental known energies, although
again the computed width (0.70 MeV) is clearly larger than the experimental one.
Moving this resonance down by use of an attractive three-body potential it is also possible
to place the resonance at 0.89 MeV, but again the computed width of 0.32 MeV is clearly
larger than the experimental one, now a factor of three larger. The corresponding full
spectrum is shown by the closed circles in the figure.

\begin{figure}
\begin{center}
\epsfig{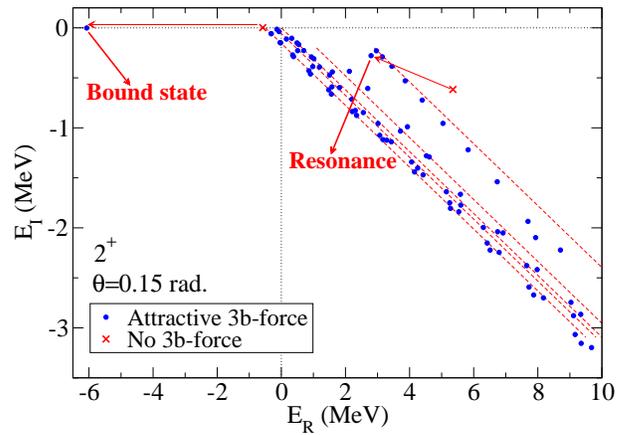}
\end{center}
\caption{(Color online) the same as Fig.\ref{fig3} for the 2$^+$ spectrum
   in $^{12}$Be ($^{10}$Be+n+n). The down sloping lines are the same thresholds as
  described in Fig.\ref{fig3}, although now the rotation angle is 0.15 rad. The
  points are the discrete states computed in the continuum. Again, 
  the crosses are the bound state and the resonance
  obtained without inclusion of any three-body force, and the closed circles have been
  obtained with an attractive three-body force which moves the bound state
  and the resonance as indicated by arrows.}
\label{fig5}
\end{figure}

In Fig.\ref{fig5} we show the $2^{+}$ complex energy spectrum which
gives rise to one bound state.  Again, we notice that both the
potentials in Fig.\ref{fig1} and the present spectrum are more
complicated than those of the $0^{-}$ states.  The properties of the
spectrum are again qualitatively very similar to those of the $0^{+}$ and
$1^{-}$ states. The resonance position without three-body potential is
now around $(5.36,-0.60)$~MeV (crosses in the figure). However, in this case the 
attractive three-body 
potential now has to be exceedingly strong to move the position down to $0.9$~MeV.  
For this reason, our calculation excludes the assignment of the spin and parity $2^+$ 
for the resonance experimentally known to be around this energy. Even more, a
moderate strength would place the resonance not much lower than about 2.8 MeV, and
having a width of 0.7 MeV. This is the spectrum shown by the solid circles in the figure.
This computed resonance could at most be compatible with 
the second reported measured resonance of the same energy with a much smaller width of
about $0.086$~MeV \cite{for94}, but more likely it would be more suitable for a higher lying
resonance, a possible candidate were observed at $5.13$~MeV by Bohlen et al. \cite{boh08}.

\begin{table}
\caption{Computed energies and widths (both in MeV) for different angular momenta and parity 
states in $^{12}$Be. The second column is the result without inclusion of a three-body
force. In the third and fourth columns a three-body force has been included to
fit the computed energy to the ones known experimentally \cite{for94,boh08}.}
\label{tab0}
\begin{tabular}{c|c|c|c}
      &  No 3b-force  &  with 3b-force & with 3b-force \\ \hline
      &  $(E_R, \Gamma)$  &  $(E_R, \Gamma)$  & $(E_R,\Gamma)$ \\ \hline
$1^-$ &  $(2.99, 0.68)$     &  $(0.89, 0.22)$  &  $(2.03, 0.50)$ \\
$0^+$ &  $(1.97, 0.70)$     &  $(0.89, 0.32)$  &  $(2.03, 0.84)$    \\
$2^+$ &  $(5.36, 1.20)$     &  $(2.76, 0.66)$  &  $(5.13, 1.13)$\\
\end{tabular}
\end{table}

A summary with the computed energies and widths for the different resonances is given in 
table~\ref{tab0}.
The second column is the result obtained without inclusion of any three-body force. In the 
third and fourth columns a three-body interaction has been included in such a way that
the energy of the computed resonances is moved down to the experimental values
given in \cite{for94} ($(E_R,\Gamma)=(0.89,0.11)$ MeV and $(E_R,\Gamma)=(2.03,0.09)$ MeV) and
in \cite{boh08}.

As we can immediately observe, once the energy of the resonances has been fitted to the experimental 
values, the corresponding widths are systematically bigger than the experimental ones. This
comparatively much larger theoretical width suggests that more than half of the structure of 
the resonances is beyond the structure of an inert core and two surrounding neutrons.

Also, the computed $0^+$ and $1^-$ states could both correspond to the two states known 
experimentally. With the help of an attractive three-body force the energy of the two 
resonances can be fitted either to 0.9 MeV or 2.0 MeV (for the $0^+$ case 
the energy of 2.0 MeV is obtained without three-body force). It is then difficult from 
the calculation to determine which of the two resonances corresponds to the $0^+$ state 
and which to the $1^-$ state. However, since the calculation without three-body force 
naturally places the $0^+$ state lower than the $1^-$ state, one could think that the 
lower resonance at 0.9~MeV is more likely of $0^+$ character,
and the one a 2.0 MeV of $1^-$ character, although the opposite can certainly not be excluded. 

In any case, from the calculation we can exclude the quantum numbers $2^+$ for the resonance at 
0.9 MeV. Even for the lowest computed 2$^+$ it is not easy to reach the energy of 2.0 MeV, and 
an energy of 2.7
MeV is the lowest we can get. This is interesting because in Ref.\cite{for94} they suggest spin
and parity 2$^+$ for the resonance at 0.9 MeV, and $2^+$, $3^-$, or $4^+$ for the one at 2.0 MeV.

Finally, the $0^{-}$ state, unless missed in the searches for bound states, can only appear 
either as a virtual state or a very broad resonance structure.

\section{Resonance structures}

The structure of the resonance wave functions is reflected in the
decomposition into partial waves of the two-body subsystems.  The
interest here is two-fold, that is bulk structure and asymptotic
large-distance configurations.  The bulk structure reflects where the
largest probability is found, which has to be at relatively small
distances since the wave function in the rotated frame vanishes
exponentially at large distance as a bound state. In an intuitive
picture of a reaction populating such a resonance, this short-distance
large-probability part would be essential for the population cross
section. On the other hand, the large-distance properties reveal which 
decay channel is preferred and in general give the branching ratios of these 
decay modes.

\subsection{Partial waves of Bulk structure}

\begin{table}
\caption{Components and weight of each component in the $1^{-}$
  resonance wave function (normalized to 1 in the complex scaling
  sense).  The energy is chosen to be $0.89$~MeV.  The upper part is the
  first Jacobi set ($x$ coordinate from neutron to neutron), and the lower 
  part corresponds to the second and third Jacobi sets ($x$ coordinate
  from core to neutron).  Only the components contributing by at least by
  1\% are included.  The orbital angular momenta are denoted by
  $\ell_x$ (between two of the particles) and $\ell_y$ (between the
  third particle and the center of mass of the other two
  particles). They couple to $L$. Correspondingly are the spins denoted
  $s_x$, $s_y$ (the spin of the third particle, not given here) and
  the total spin, $S$, obtained by coupling.  The value of $K_{max}$
  gives the maximum value for the hypermomentum \cite{nie01} employed for this
  component. The probability in \% is in the last column, and it is denoted as
  weight.  }
\begin{center}
\begin{tabular}{cccccc|c}
$\ell_x$ & $\ell_y$ & $L$ & $s_x$ & $S$ & $K_{max}$ & Weight (\%)  \\
\hline         
   0  &  1  &  1  &  0  &  0 & 300 &   29.7  \\
   1  &  0  &  1  &  1  &  1 & 200 &   10.8  \\
   1  &  2  &  1  &  1  &  1 & 160 &   1.5  \\
   1  &  2  &  2  &  1  &  1 & 300 &   34.3  \\
   2  &  1  &  1  &  0  &  0 & 200 &   6.8  \\
   2  &  3  &  1  &  0  &  0 &  60 &   3.2  \\
   3  &  2  &  1  &  1  &  1 &  60 &   1.7  \\
   3  &  2  &  2  &  1  &  1 &  60 &   5.6  \\ 
   3  &  4  &  2  &  1  &  1 &  40 &   4.9  \\
   4  &  3  &  2  &  0  &  0 &  40 &   1.0  \\ \hline
   0  &  1  &  1  & 1/2 &  1 & 160 &   1.1  \\
   1  &  2  &  1  & 1/2 &  0 & 200 &  18.4  \\
   1  &  2  &  1  & 1/2 &  1 & 120 &   6.4  \\
   1  &  2  &  2  & 1/2 &  1 & 300 &  20.9  \\
   2  &  1  &  1  & 1/2 &  0 & 300 &  20.3  \\
   2  &  1  &  1  & 1/2 &  1 & 200 &   6.2  \\
   2  &  1  &  2  & 1/2 &  1 & 300 &  25.1  \\
\end{tabular}
\end{center}
\label{tab1}
\end{table}

We begin with the bulk structure of the partial-wave decomposed
resonance wave function.  We show in table~\ref{tab1} the
probabilities of finding different configurations within the $1^{-}$
resonance.  The energy is adjusted with the three-body potential to be
$0.89$~MeV.  The partial-wave decompositions do not change substantially by
increasing this energy up to $2.0$~MeV.  The dominating configurations are
total orbital angular momentum $L=1,2$.  The $L=1$ ($2$)
configuration is with either $p$ or $s$ ($d$)-waves between the two
neutrons, combined respectively with $s$ ($d$) or $p$-waves of their
center of mass around the core.  The Pauli principle then determines
the spin combinations producing either $0$ or $1$.  Expressing the
same wave function in the other Jacobi system we find four comparable
components with neutron-core in $p$ or $d$ combined with $d$ or $p$
partial waves of the last neutron around the neutron-core
center of mass.

\begin{table}
\caption{Components and weight of each component in the $0^{+}$
  resonance wave function at $0.89$~MeV. The notation is as in table~\ref{tab1}. }
\begin{center}
\begin{tabular}{cccccc|c}
$\ell_x$ & $\ell_y$ & $L$ & $s_x$ & $S$ & $K_{max}$ & Weight (\%)  \\
\hline         
   0  &  0  &  0  &  0  &  0 & 240 &   61.6  \\
   1  &  1  &  1  &  1  &  1 & 240 &   28.3  \\
   2  &  2  &  0  &  0  &  0 & 200 &   10.1  \\ \hline
   0  &  0  &  0  & 1/2 &  0 & 100 &   3.2  \\
   1  &  1  &  0  & 1/2 &  0 & 100 &   1.4  \\
   1  &  1  &  1  & 1/2 &  1 & 100 &   5.6  \\
   2  &  2  &  0  & 1/2 &  0 & 200 &  65.0  \\
   2  &  2  &  1  & 1/2 &  1 & 200 &  24.8  \\
\end{tabular}
\end{center}
\label{tab2}
\end{table}

The $0^{+}$ resonance is also a suitable candidate for both the two
lowest-lying observed resonances. We show in table~\ref{tab2} its bulk
structure for an energy of $0.89$~MeV. The weights of these
configurations would only change very little by increasing the energy
to about $2.0$~MeV.  The two neutrons are in relative $s$, $p$, or $d$
waves and strongly decreasing probability with orbital angular
momentum. This corresponds to the neutron-core essentially entirely in
$d$-waves combined with the last neutron in $d$-waves.

\begin{table}
\caption{Components and weight of each component in the $2^{+}$
  resonance wave function at 5.36~MeV. The notation is as in table~\ref{tab1}. }
\begin{center}
\begin{tabular}{cccccc|c}
$\ell_x$ & $\ell_y$ & $L$ & $s_x$ & $S$ & $K_{max}$ & Weight (\%)  \\
\hline         
   0  &  2  &  2  &  0  &  0 & 350 &  11.6  \\
   1  &  1  &  1  &  1  &  1 & 450 &  32.5  \\
   1  &  1  &  2  &  1  &  1 & 100 &   1.8  \\ 
   1  &  3  &  2  &  1  &  1 & 100 &   2.6  \\ 
   2  &  0  &  2  &  0  &  0 & 350 &   7.9  \\ 
   2  &  2  &  2  &  0  &  0 & 350 &  18.8  \\ 
   2  &  4  &  2  &  0  &  0 & 100 &   1.5  \\ 
   3  &  1  &  2  &  1  &  1 & 100 &   2.9  \\ 
   3  &  3  &  1  &  1  &  1 & 100 &   8.7  \\ 
   3  &  3  &  2  &  1  &  1 & 100 &   5.9  \\ 
   4  &  4  &  2  &  0  &  0 & 100 &   4.2  \\ \hline
   1  &  1  &  1  & 1/2 &  1 & 200 &   1.1  \\
   2  &  0  &  2  & 1/2 &  0 & 500 &   1.1  \\
   2  &  2  &  1  & 1/2 &  1 & 500 &  36.6  \\
   2  &  2  &  2  & 1/2 &  0 & 500 &  43.1  \\
   2  &  2  &  2  & 1/2 &  1 & 300 &  15.6  \\
\end{tabular}
\end{center}
\label{tab3}
\end{table}

The $2^{+}$ resonance is not a natural candidate for the lowest
resonance at $0.89$ but very suitable for a high-lying resonance at
about $5$~MeV, and may be also for the $2.03$~MeV resonance.
We show in table~\ref{tab3} its bulk structure for an energy of
$5.4$~MeV.  The weights of these configurations are rather
insensitive to variations of the energy by several MeV.  The two
neutrons are, as for $0^{+}$, in relative $s$, $p$, or $d$ waves but
now the largest probability is for $p$-waves. The last neutron is
correspondingly in $d$, $p$, or $s$ waves.  In the other Jacobi system
this corresponds to the neutron-core essentially entirely in $d$-waves
combined with the last neutron in $d$-waves around this structure of
the $^{11}$Be system.

\begin{table}
\caption{Components and weight of each component in the $0^{-}$ bound
  state wave function with an energy of $-0.518$ MeV, slightly more bound than the
  $-0.504$ MeV corresponding to the lowest one-body threshold, i.e. the $1/2^+$ ground state
  of $^{11}$Be. The notation is as in table~\ref{tab1}. }
\begin{center}
\begin{tabular}{cccccc|c}
$\ell_x$ & $\ell_y$ & $L$ & $s_x$ & $S$ & $K_{max}$ & Weight (\%)  \\
\hline         
   1  &  0  &  1  &  1  &  1 & 500 &  82.8  \\
   1  &  2  &  1  &  1  &  1 & 400 &   3.7  \\
   3  &  2  &  1  &  1  &  1 & 200 &   8.9  \\
   3  &  4  &  1  &  1  &  1 & 100 &   2.1  \\
   5  &  4  &  1  &  1  &  1 &  40 &   1.0  \\ \hline
   0  &  1  &  1  & 1/2 &  1 & 500 &  54.4  \\
   1  &  0  &  1  & 1/2 &  1 & 500 &  44.4  \\
\end{tabular}
\end{center}
\label{tab4a}
\end{table}

The last of the appropriate quantum numbers is $0^{-}$ which
disappears into the continuum as soon as it is attempted lifted a
little above the energy of the $^{11}$Be ground state.  It is not
possible to place such a resonance structure closer to the threshold
of the excited state of $^{11}$Be, and much less in the true
three-body continuum of $^{12}$Be. When the $0^-$ state is genuinely bound,
with a binding energy of about $-1$ MeV, its bulk structure 
is given by table V of Ref.\cite{rom07b}. This structure does not
change very much with the binding energy, and its main characteristics
are similar to the ones found when the system is made to be bound, but just
below the $-0.504$ MeV corresponding to the lowest one-body threshold, i.e.
the $1/2^+$ ground state of $^{11}$Be. This is shown in table~\ref{tab4a}
for a three-body energy of $-0.518$ MeV. As in Ref.\cite{rom07b}, the two neutrons 
in $p$-waves dominates with the core in an $s$-wave. The neutron-core structure is
roughly equally distributed in $s$ and $p$-waves with $p$ and $s$
partial waves for the last neutron in the motion around this $^{11}$Be
state.

\begin{table}
\caption{Components and weight of each component in the $0^{-}$
  resonance wave function with an energy of $-0.479$ MeV, which is just
  above the $-0.504$ MeV corresponding to the lowest one-body threshold, i.e. 
  the $1/2^+$ ground state
  of $^{11}$Be.  The notation is as in table~\ref{tab1}. }
\begin{center}
\begin{tabular}{cccccc|c}
$\ell_x$ & $\ell_y$ & $L$ & $s_x$ & $S$ & $K_{max}$ & Weight (\%)  \\
\hline         
   1  &  0  &  1  &  1  &  1 & 500 &   9.2  \\
   1  &  2  &  1  &  1  &  1 & 400 &  12.7  \\
   3  &  2  &  1  &  1  &  1 & 200 &  14.7  \\
   3  &  4  &  1  &  1  &  1 & 200 &  13.5  \\
   5  &  4  &  1  &  1  &  1 & 100 &  12.6  \\
   5  &  6  &  1  &  1  &  1 & 100 &  10.4  \\
   7  &  6  &  1  &  1  &  1 &  60 &   9.0  \\
   7  &  8  &  1  &  1  &  1 &  40 &   7.1  \\ 
   9  &  8  &  1  &  1  &  1 &  40 &   6.0  \\
   9  & 10  &  1  &  1  &  1 &  40 &   4.7  \\ \hline
   0  &  1  &  1  & 1/2 &  1 & 500 &  55.5  \\
   1  &  0  &  1  & 1/2 &  1 & 500 &  25.3  \\
   1  &  2  &  1  & 1/2 &  1 & 400 &   2.4  \\
   2  &  1  &  1  & 1/2 &  1 & 400 &  10.4  \\
   2  &  3  &  1  & 1/2 &  1 & 200 &   1.0  \\
   3  &  2  &  1  & 1/2 &  1 & 200 &   3.5  \\
\end{tabular}
\end{center}
\label{tab4b}
\end{table}

However, when increasing the energy of the $0^{-}$ bound state, and lifting it above
the $^{11}$Be ground state threshold, the partial wave
decomposition changes.  We show in table~\ref{tab4b} the results for a
three-body energy of $-0.479$ MeV.  The neutron-neutron $p$-wave is now distributed 
over many components of
higher relative angular momenta with correspondingly higher center of
mass angular momenta around the core.  In the second Jacobi system we
find that the neutron-core $s$-wave component basically remains
unchanged whereas the $p$-wave becomes distributed over many partial
waves. Therefore, above the threshold the configuration is about $50\%$
of the neutron-$^{10}$Be in the ground state of $^{11}$Be. The other
half of the probability is distributed among many partial waves. This fact
strongly suggests that this part of the resonance structure approaches
a free plane wave solution. Thus, the $0^{-}$ structure above the $^{11}$Be
ground state threshold resembles a coherent combination of a two-body
structure, neutron-$^{11}$Be($1/2^+$), and a three-body continuum
state without resonance structure. 

The energy ordered sequence of resonances is then
most consistently in the model given by $0^+$, 1$^-$, and 2$^+$, while
$0^-$ (unless missed in the searches for bound states) only appears as
an attractive potential related to a virtual state.

\subsection{Spatial properties of the resonances}

Resonances are quasi-stable systems produced by the presence of some potential
barrier the keeps the particles close to each other for a certain amount of time. 
Eventually the particles tunnel through the barrier and the resonance decays. Even 
if this fact implies a non-square integrable wave function for the system, it reasonable
to ask ourselves about the spatial distribution of the resonance constituents while kept together
inside the potential pocket. This analysis can be easily made through the complex scaled
resonance wave function. In this way, we get rid of the divergent part of the wave function, which 
is associated to the spatial distribution after decay, while the structure of the inner part  
is maintained.

In particular, we shall construct the spatial distribution function by integrating the square 
of the wave function with respect to the directions of the $\bd{x}$ and $\bd{y}$ Jacobi coordinates.
More precisely, we define it as:
\begin{equation}
D(J^\pi;r_x,r_y)=\int r_x^2 r_y^2 (\Psi^{J^\pi}(\bd{r}_x,\bd{r}_y))^2 d\Omega_x d\Omega_y,
\label{distr}
\end{equation}
where $r_x$ is the distance between the two particles connected by the Jacobi coordinate $\bd{x}$,
and $r_y$ is the distance between the third particle and the center of mass of the first two. 
Obviously, for a system like $^{12}$Be, with a core and two neutrons, we can construct two spatial
distribution functions, since $r_x$ can refer either to the distance between the two neutrons or
to the distance between one of the neutrons and the core. 

It is important to note that the wave
function $\Psi^{J^\pi}$ is complex, and therefore the spatial distribution function $D$ is complex
as well (the square of the wave function is not the square of the modulus \cite{moi98}). However,
as discussed in \cite{moi98}, the imaginary part of the computed observables can be interpreted
as the uncertainty of the measuring computed observable, while the real part is associated to the
value of the observable itself (this is what happens for instance with the expectation value
of the hamiltonian, which is the complex energy of the resonance). For this reason, we shall
in the future consider only the real part of the distribution function $D$ defined in Eq.(\ref{distr}).

Concerning the $1^-$ states, the bound state is mainly built on the lowest and broadest 
adiabatic potential in Fig.\ref{fig1}. This potential does not show any barrier, and therefore
it can not be responsible for the appearance of $1^-$ resonances. This is not happening with
the second attractive potential, which together with the potential pocket exhibits as well
a barrier providing the stability necessary for a finite width. This is in fact the potential 
responsible for the appearance of the $1^{-}$ resonance whether it is placed at $0.89$ or
at $2.03$~MeV. 

\begin{figure}
\begin{center}
\epsfig{file=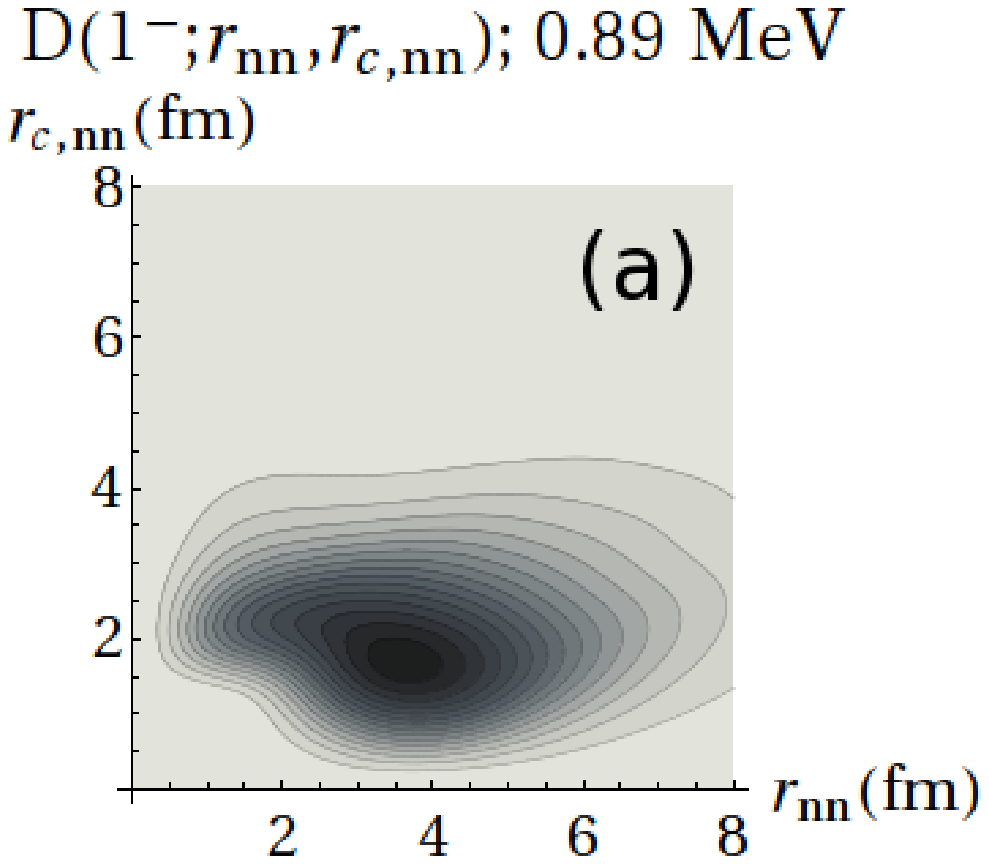,width=4.2cm,angle=0}
\epsfig{file=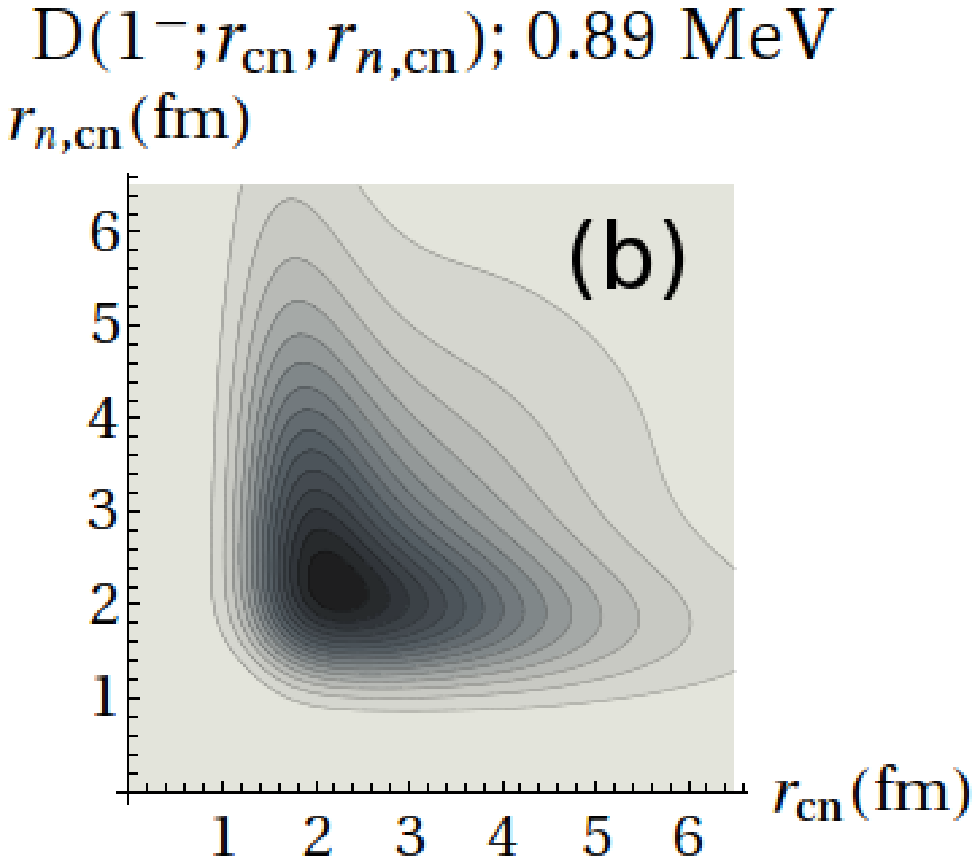,width=4.2cm,angle=0}
\epsfig{file=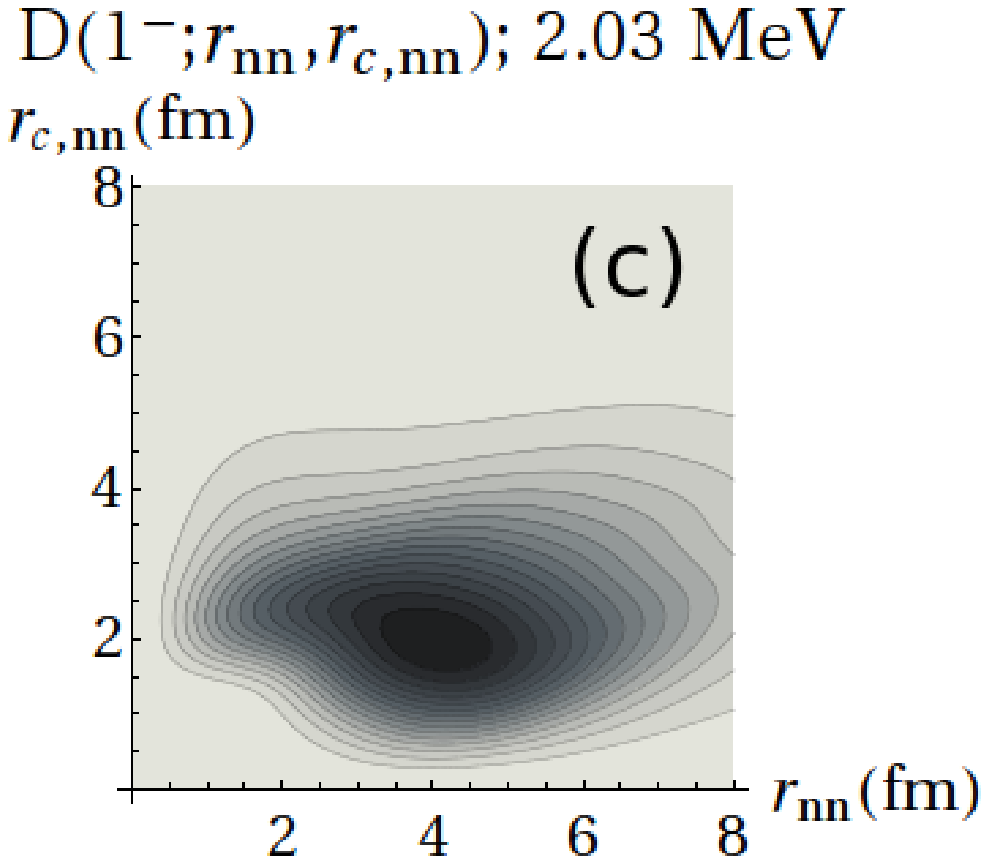,width=4.2cm,angle=0}
\epsfig{file=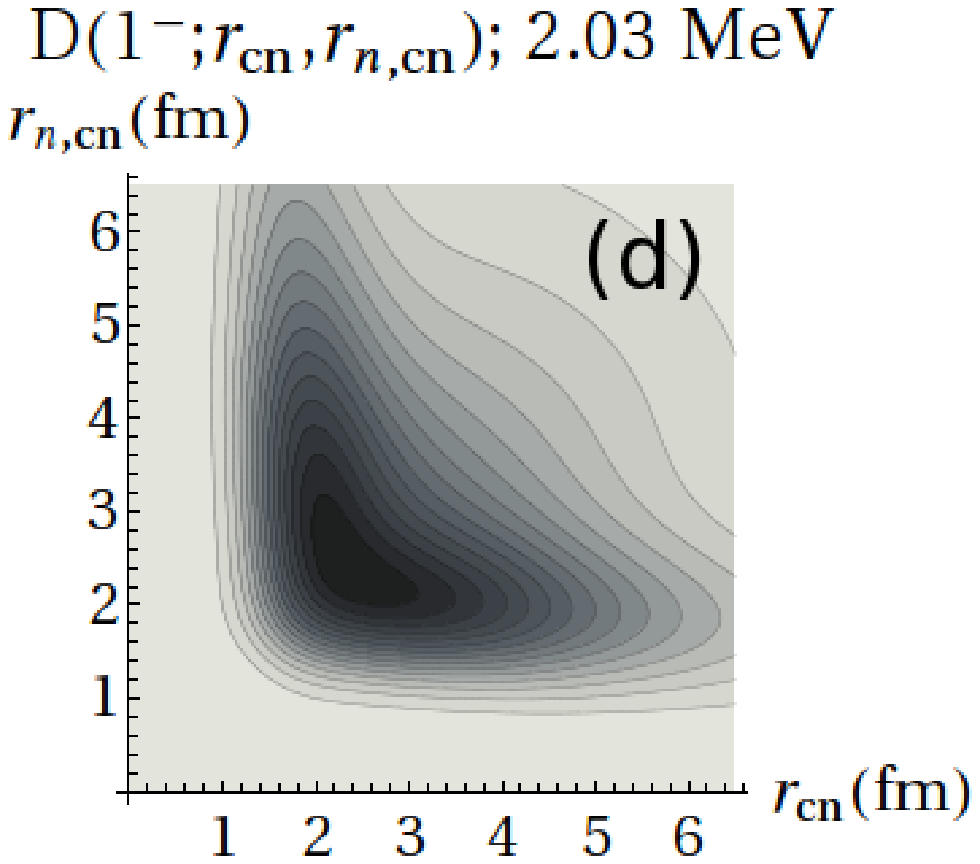,width=4.2cm,angle=0}
\end{center}
\caption{(Color online) Contour plot of the spatial distribution function in Eq.(\ref{distr}) 
for the $1^-$ resonance in $^{12}$Be when placed at 0.89 MeV (upper part), and when placed at 
2.03 MeV (lower part). The figures on the left correspond to the first Jacobi set ($r_x=r_{nn}$ 
is the distance between the two neutrons), and the figures on the right correspond to the second 
and third Jacobi sets ($r_x=r_{cn}$ is the distance between one of the neutrons and the $^{10}$Be core).
The darker the color in the plot the higher the value of the distribution function.}
\label{fig1-}
\end{figure}

Making use now of Eq.(\ref{distr}) we can investigate the spatial distribution of the two
neutrons and the $^{10}$Be core in the $1^-$ resonance. This is done in Fig.~\ref{fig1-},
where we show the contour plot of the spatial distribution function for the two
resonances energies considered in this work. In the left part of the figure we give the 
results in the first Jacobi set, where the horizontal and vertical axes represent the distance
between the two neutrons ($r_{nn}$) and the distance between the core and the center of mass of
the two neutrons ($r_{c,nn}$), respectively. In the right part of figure the second and third
Jacobi sets are used, meaning that the coordinates used are the distance between the
core and one of the neutrons ($r_{cn}$), and the distance between the second neutron and the center
of mass of the $n-^{10}$Be system ($r_{n,cn}$). Obviously, a darker color in the figure implies
a higher value of the distribution $D$.

As we can see, there are no relevant differences between the spatial structure depending on the
energy. The increase in energy from about 1 MeV to about 2 MeV only moves very little the maxima 
in the figure towards bigger separation between particles, but the structure of the system
remains the same. In the two cases shown, with energies 0.89 MeV (upper part) and 2.03 MeV (lower part),
the system shows a clean maximum for the two neutrons about 4 fm far apart from each other, and the core
about 2 fm from the center of mass of the two neutrons. This is essentially an isosceles triangle with the
two neutrons defining the different side ($\sim 4$ fm long), while the two other sides are 
about 2.7 fm long, which  corresponds to the distance between the core and each of the two neutrons.
This geometry is consistent with the distribution seen in the right part of the figure, where the maximum
is found for $r_{cn}\approx 2.6-2.7$ fm (due to the large mass of the core compared to the neutron mass
we have that $r_{n,cn} \approx r_{cn}$).

\begin{figure}
\begin{center}
\epsfig{file=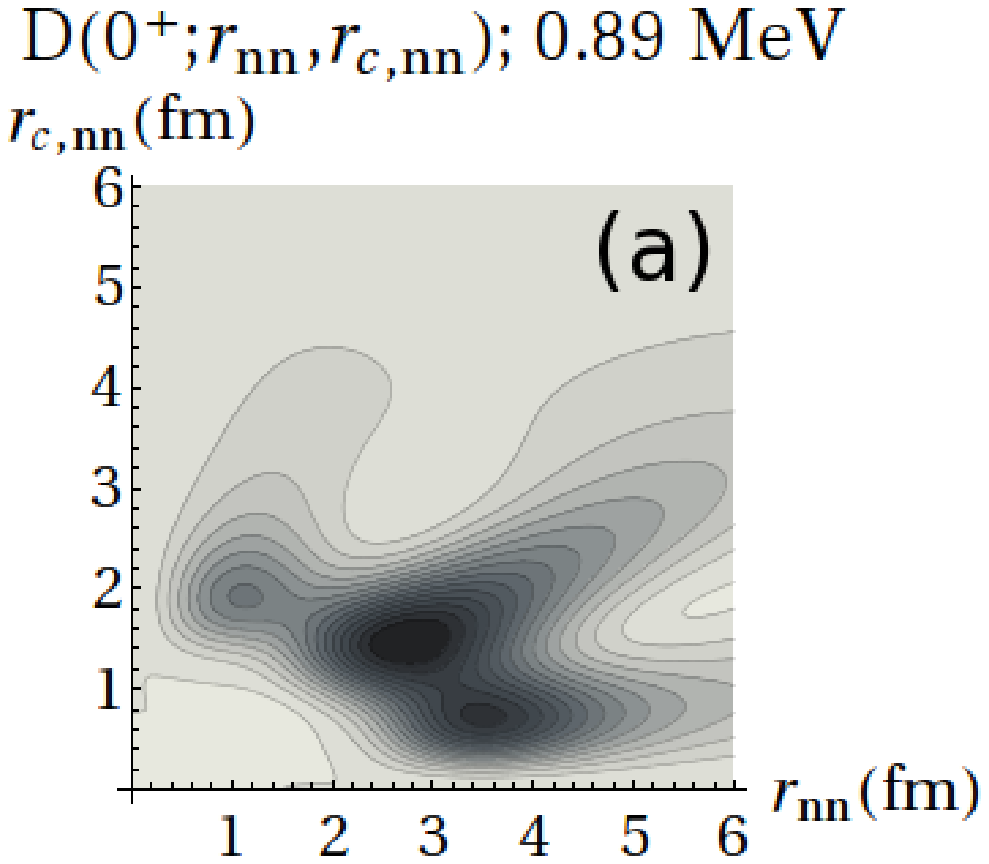,width=4.2cm,angle=0}
\epsfig{file=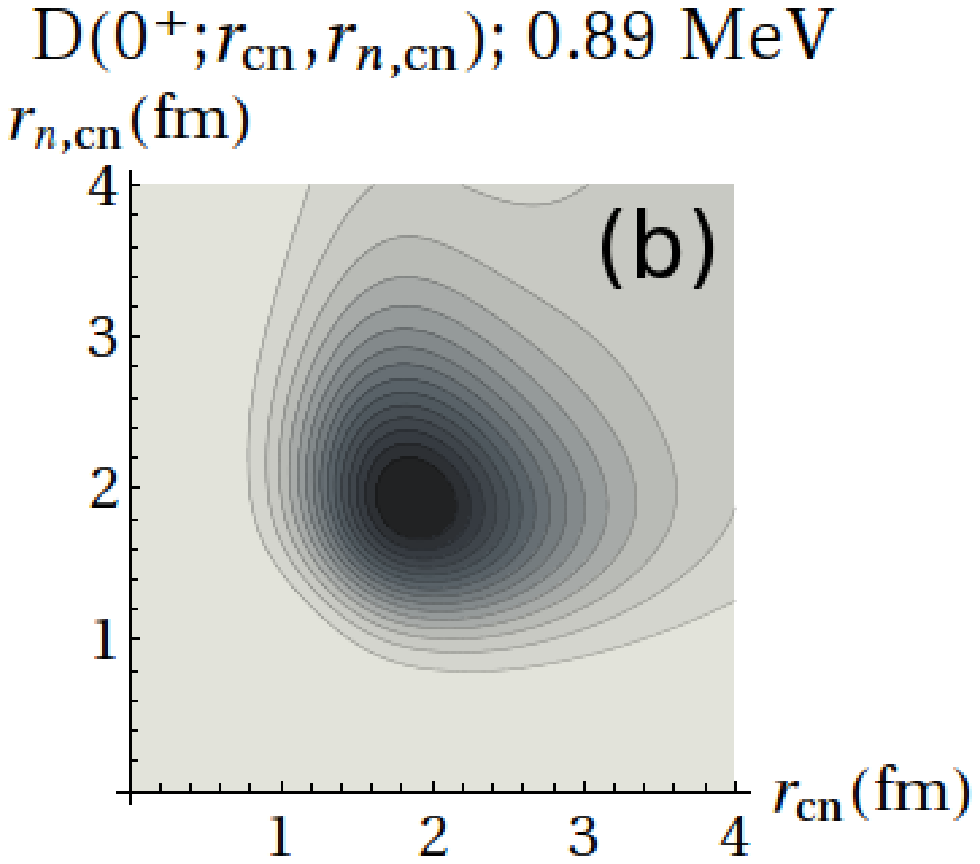,width=4.2cm,angle=0}
\epsfig{file=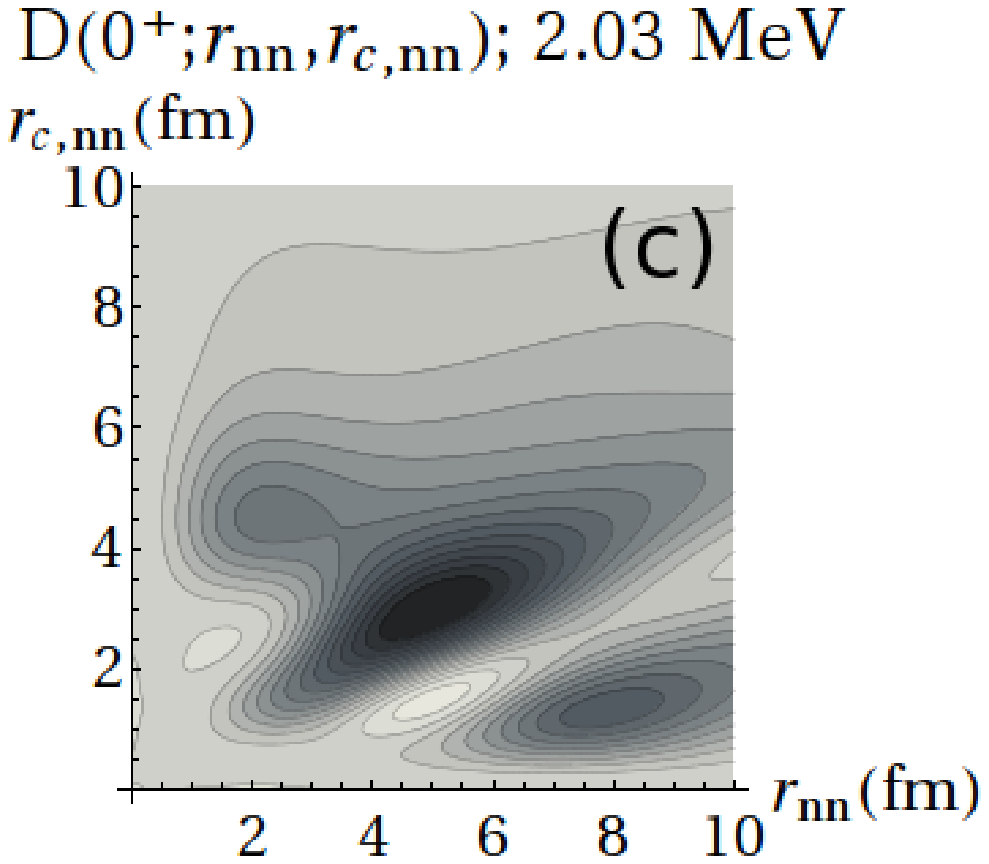,width=4.2cm,angle=0}
\epsfig{file=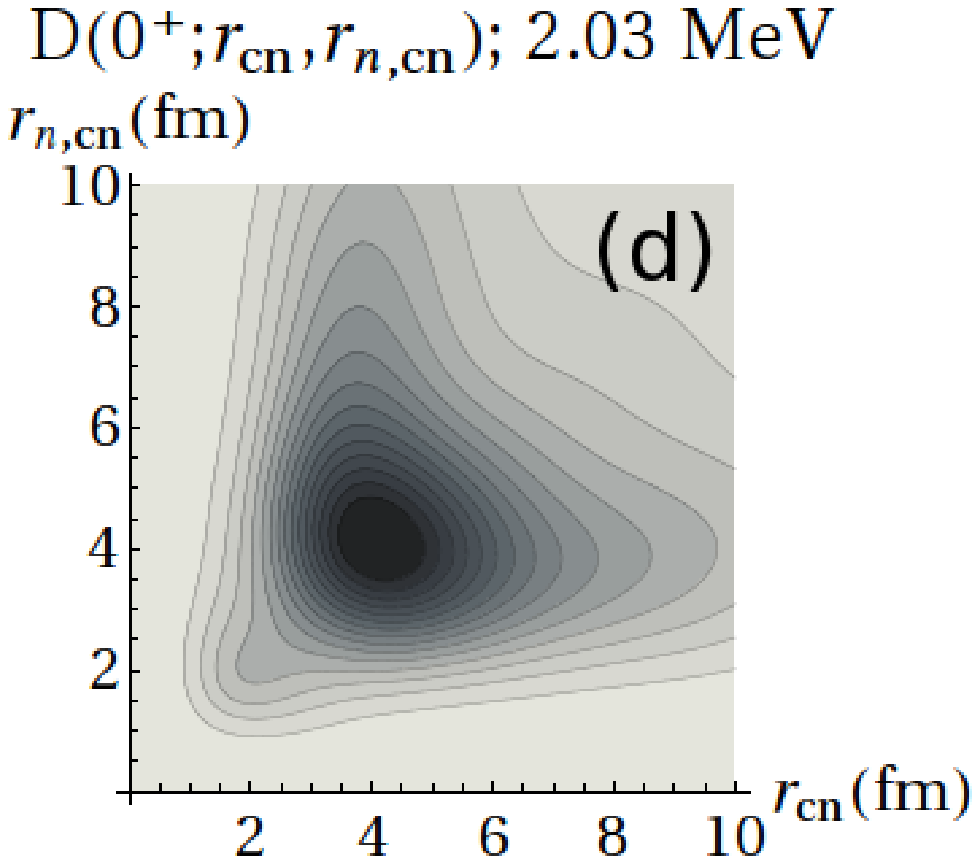,width=4.2cm,angle=0}
\end{center}
\caption{(Color online) The same as Fig.~\ref{fig1-} for the 0$^+$ resonance in $^{12}$Be when
placed at 0.89 MeV (upper part), and when placed at 2.03 MeV (lower part).}
\label{fig0+}
\end{figure}

The $0^{+}$ structures are found in two bound as well as in one
resonance state. The bound states are built on the two attractive
adiabatic potentials in Fig.\ref{fig1} without any barrier, see
\cite{rom07b}.  The $0^{+}$ resonance is related to the other two
attractive potentials in Fig.\ref{fig1}, both of them showing a 
barrier.  The relative weight for the
energy of $0.89$~MeV is $66\%$ on the deepest and narrowest potential
and the remaining $44\%$ is from the high-lying attractive potential
with a barrier.  Increasing the energy to $2.03$~MeV, the latter
high-lying potential becomes responsible for $92\%$ whereas the
remaining 8\% is shared among the many other potentials.  The structure has
changed substantially away from that of the deep and narrow potential.

The spatial distribution for the $0^+$ resonance is shown in Fig.~\ref{fig0+}.
In this case the jump from 0.89 MeV to 2.03 MeV increases the separation between particles
more significantly than for the $1^-$ resonance. The maximum of the distribution appears
for a distance between neutrons smaller than 3 fm when the energy is 0.89 MeV (upper part), 
and about 5 fm when the energy is 2.03 MeV (lower part). In any case, the spatial
distribution is similar (but scaled) in both cases. In fact, in both cases we observe
two additional maxima corresponding to an almost aligned distribution with the $^{10}$Be
core very close to the center of mass of the two neutrons, an another corresponding to a dineutron
(the two neutrons close to each other) and the core far apart.

The $2^{+}$ structure is also found as a bound state which is
predominantly  built on the lowest attractive adiabatic potential.
The additional $2^{+}$ resonance structure which occurs ``naturally''
without a three-body potential at $5.36$~MeV is almost entirely built
on the second potential with an attractive pocket at small distance.
Moving this resonance down with an attractive three-body potential
increases the contribution from the potential with the attractive
pocket at about $5$~fm in Fig.\ref{fig1}.  For an energy of $2.77$~MeV
these two potentials share roughly equally the probabilities.

\begin{figure}
\begin{center}
\epsfig{file=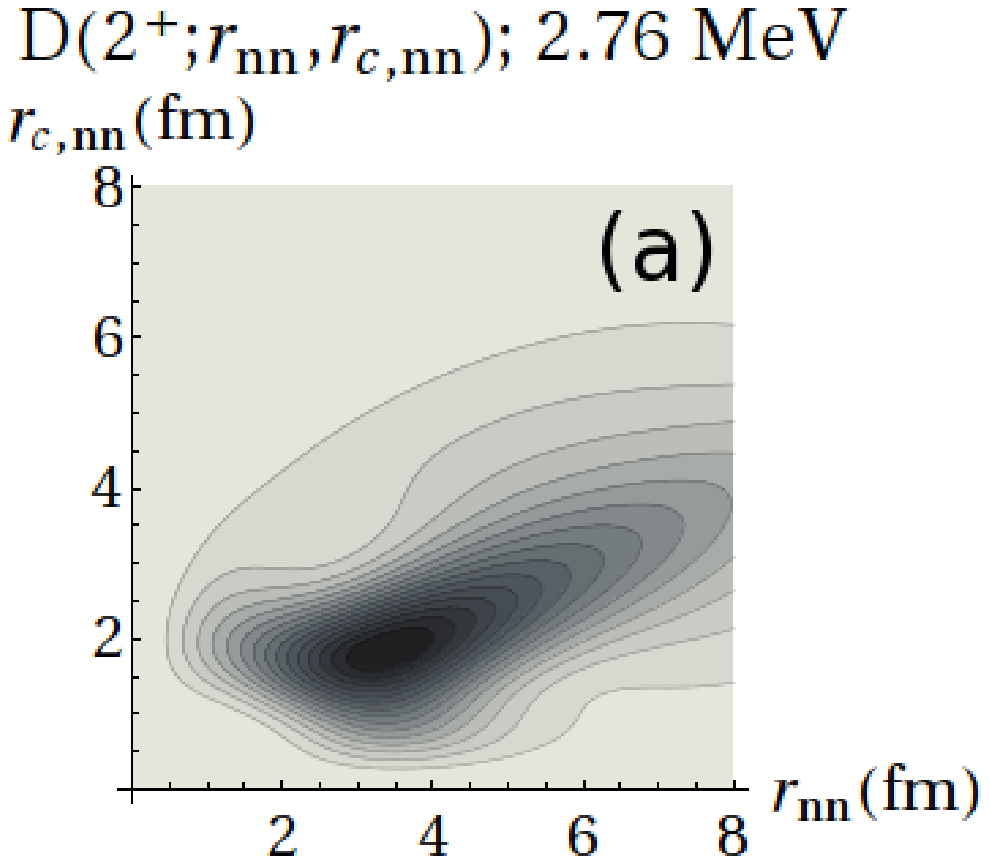,width=4.2cm,angle=0}
\epsfig{file=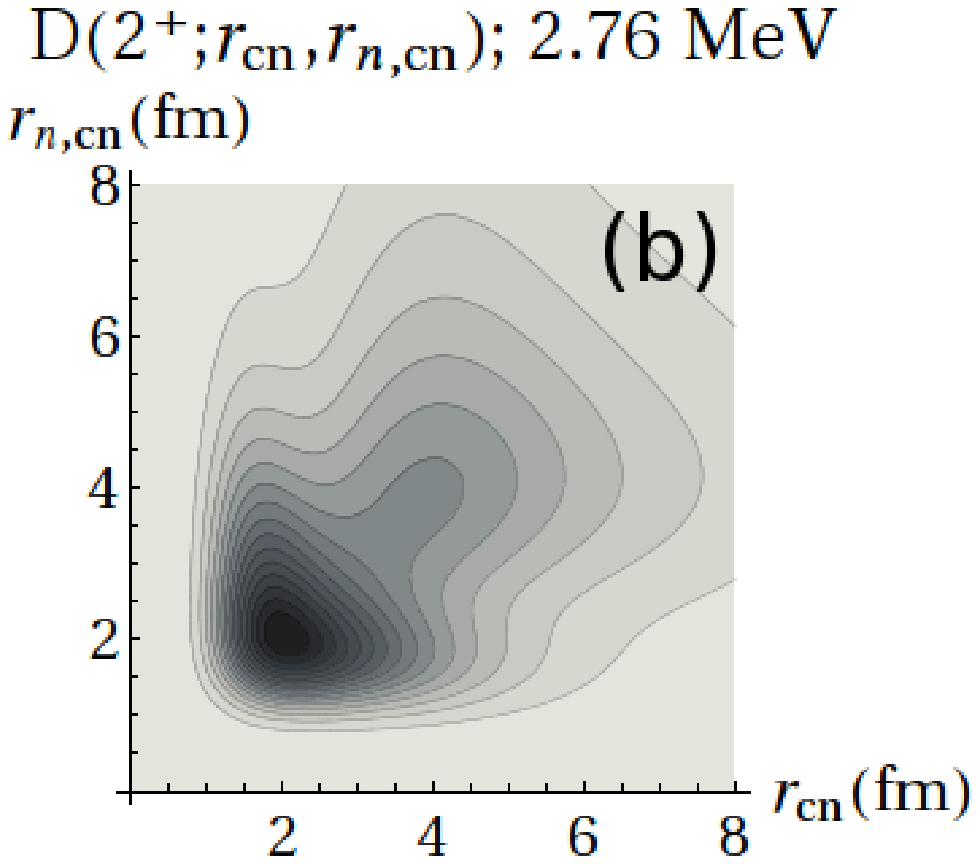,width=4.2cm,angle=0}
\epsfig{file=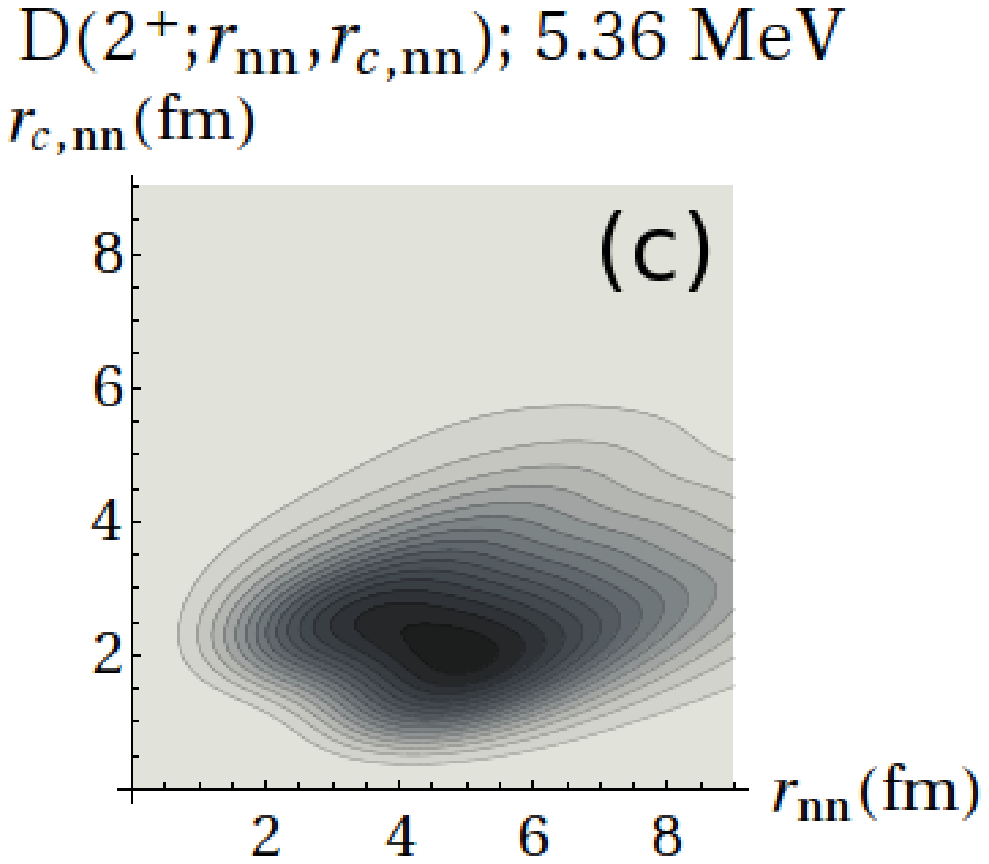,width=4.2cm,angle=0}
\epsfig{file=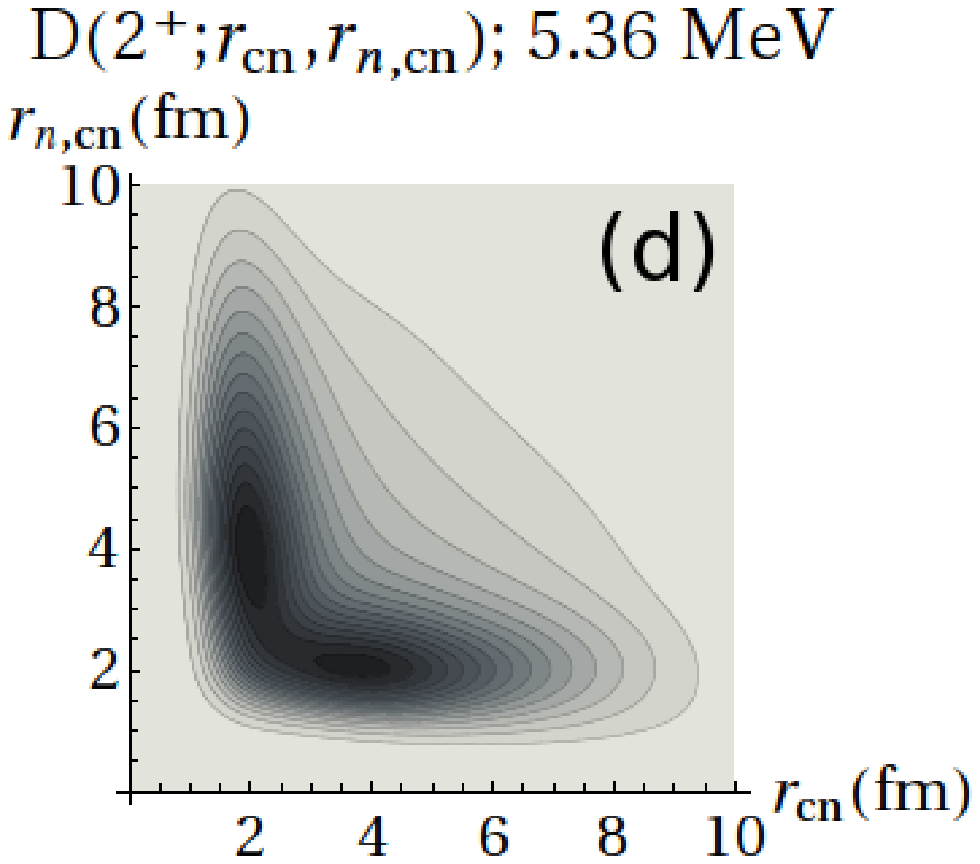,width=4.2cm,angle=0}
\end{center}
\caption{(Color online) The same as Fig.~\ref{fig1-} for the 2$^+$ resonance in $^{12}$Be when
placed at 2.76 MeV (upper part), and when placed at 5.36 MeV (lower part).}
\label{fig2+}
\end{figure}

The spatial distribution function for the $2^+$ resonance is shown in
Fig.~\ref{fig2+}. Contrary to what happened in the $1^-$ and $2^+$ cases, 
where the energy change between the two cases shown was relatively small,
now the energy variation is bigger (from 2.76 MeV to 5.36 MeV), and the
upper and lower parts in the figure show clear differences.
This is more easily seen in the second and third Jacobi sets (right part
of the figure). When the energy is chosen to be 2.76 MeV the two neutrons like
more to be equally separated from the core (a bit more than 2 fm), but for an
energy of 5.36 MeV a structure with a neutron close to the core (about 2 fm) and
the second one far from it (around 4.5 fm) is preferred.

\section{Decay and production of $^{12}$Be}

The calculated structures can be used to determine both the probability of the 
different decay modes and the probability of populating a resonance. Both the
population and decay of the resonances in $^{12}$Be can be studied experimentally.
The two quantities are very dependent on the quantum numbers of the resonances, 
hence a comparison between the theoretically calculated and the experimentally 
determined strength can be used to assign quantum numbers for the resonances.

\subsection{Decay modes}

The decay channels are determined from the large-distance structure of
the radial wave functions. As seen in Fig.\ref{fig1}, each adiabatic potential
corresponds to a very specific asymptotic configuration, i.e., bound $^{11}$Be 
(either in the ground state or in the excited state) plus a neutron, $^{11}$Be populating
a two-body resonance plus a neutron, or the three constituents in the continuum.
This is because the observable final state
momenta of each of the emerging particles precisely are reflected in
the coordinate wave function at asymptotic large distances \cite{alv08,alv07}.
The probability of decaying through a given channel is then the
probability of occupying the adiabatic potential describing that 
channel at large distance.

\begin{table}
\caption{Branching ratios (in \%) after decay of the resonance with angular 
  momentum and parity $J^\pi$.  The resonance energies and widths, $E$ and 
  $\Gamma$, are given in MeV. For each decay channel we specify the half-integer
  angular momentum of $^{11}$Be or the $0^+$ angular momentum of $^{10}$Be, and
  where each $n$ indicates emission of one neutron.  } 
\begin{center}
\begin{tabular}{cc|ccccc}
$J^\pi$ & $(E,\Gamma)$ & $\frac{1}{2}^+$+n \;& $\frac{1}{2}^-$+n \;& $0^+$+n+n & \;
                         $\frac{3}{2}^+$+n \; & $\frac{5}{2}^+$+n \\ 
\hline         
   $0^+$ & (0.89,0.32)  &  8.6   &  59.5  &  31.8  &  0.0  &  0.0 \\
   $0^+$ & (2.03,0.84)  &  0.8   &  12.8  &  48.4  &  0.0  & 38.0 \\
   $1^-$ & (0.89,0.22)  &  17.4  &  74.5  &  8.1   &  0.0  &  0.0 \\
   $1^-$ & (2.03,0.50)  &  6.3   &  84.5  &  7.7   &  0.0  &  1.5 \\
   $2^+$ & (2.76,0.66)  & 14.9   &   0.9  &  65.2  &  0.0  & 19.0 \\
   $2^+$ & (5.36,1.20)  &  8.2   &   1.8  &  55.5  & 13.9  & 20.6 \\
\end{tabular}
\end{center}
\label{tab5}
\end{table}

The angular momenta and parities of the experimentally observed
resonances are not known, although in some cases their values
are suggested as the most likely according to the experimental conditions.  
The energies chosen for each of the resonances computed in this
work meet these conditions, and we therefore
calculate branching ratios for the $0^{+}$, $1^{-}$, and $2^{+}$
resonances each of them placed at two different energies.
The computed branching ratios for each of them are given
in table~\ref{tab5}, where the energy and width of each resonance 
$(E,\Gamma$) is given in MeV. The decay channels denoted by ``$\frac{1}{2}^+$+n''
and ``$\frac{1}{2}^-$+n'' represent emission of one neutron plus
$^{11}$Be either in the ground state or in its bound excited state. The 
direct decay into the three-body continuum, with the $^{10}$Be core 
in the $0^+$ ground state, is denoted by ``$0^+$+n+n''. Finally,
``$\frac{3}{2}^+$+n'' and ``$\frac{5}{2}^+$+n'' correspond to
sequential decay through the $\frac{3}{2}^+$ and $\frac{5}{2}^+$
resonances in $^{11}$Be, which eventually decay into the two-body
continuum of $^{10}$Be and a neutron.

For $0^{+}$ resonance at $E=0.89$ MeV the preferred
decay channels are either one-neutron decay to the
excited $1/2^-$ state of $^{11}$Be or direct decay to the three-body
continuum.  Only about 8.6\% of the decay takes place through the ground
state of $^{11}$Be. For the higher energy of $E=2.03$~MeV, since the energy
of the $5/2^+$ resonance in $^{11}$Be is 1.28 MeV, the probability for
sequential decay through this resonance now becomes prominent.
The direct decay is also increased at the expense of the one-neutron decay 
branch, that reduces from about 58\% to about 14\%, and where in particular 
the decay through the ground state of $^{11}$Be almost disappears. 

If the lowest resonance at about 0.9 MeV is a $1^{-}$ state, we find that most
decays proceed by one-neutron emission to the excited $^{11}$Be state
and a smaller but significant fraction decays to the $^{11}$Be ground
state. The direct decay into the continuum amounts only to about 8\%.
If the $1^{-}$ state is at about $2$~MeV the branching ratios
are even larger for decay into the excited $^{11}$Be state and less
than $10\%$ into the ground state, while the direct decay remains
around $8\%$.

The lowest resonance have been seen in a d($^{11}$Be,p)$^{12}$Be experiment
at ISOLDE, by gating on the gamma from the sequential decay through
the $1/2^-$ bound state in $^{11}$Be \cite{joh10}. This is the first experiment in which
a given decay channel from a resonance in $^{12}$Be has been singled out,
and could open up for branching ratio measurement, but the analysis is still
ongoing. A measurement of the branching ratio will give a strong indication
of the quantum numbers of the resonances.  As we can see, the $^{12}$Be 
resonance at about 0.9 MeV and the one about 2.0 MeV would have rather 
different decay branching ratios depending on which of them is the $0^+$
state and which the $1^-$ state. A 0$^+$ state would decay with a large component
of direct decay into the continuum. This component reduces drastically in the
case of the $1^-$ resonance, that clearly prefers to decay through the $1/2^-$ 
state in $^{11}$Be.

The calculated $2^{+}$ resonance is naturally placed at a higher
energy, or alternatively predominantly of a non-three body structure.
For an energy of $5.36$~MeV the dominating decay channel is direct
decay or sequential decays through either $3/2^+$ or $5/2^+$
resonances in $^{11}$Be.  For a lower energy of $2.76$~MeV the direct
decay is even more dominating, while decay to the ground state has
increased at the expense of the sequential decay through the
high-lying $3/2^+$ resonance, which for energy reasons disappears
(the energy of $3/2^+$ resonance in $^{11}$Be is 2.90 MeV).

\subsection{Production strength}

While the decay of the resonances of $^{12}$Be are still to be measured, 
the continuum of $^{12}$Be has been probed in both neutron transfer and
charge exchange experiments. So far only two experiments have 
been able to clearly identify any resonances \cite{for94,boh08}, and the information
is limited to excitation energy and width. Both experiments show clear resonances
at $0.89$~MeV and $2.03$~MeV, and tentatively quantum numbers of $2^+$ and $0^+$ or $4^+$ 
is suggested from DWBA calculations. A large peak at $5.13$~MeV seen in the 
$^9$Be($^{12}$C,$^{9}$C)$^{12}$Be experiment \cite{boh08} strongly indicates of a 
third resonance. Unfortunately this resonance cannot be confirmed by the $^{10}$Be(t,p)$^{12}$Be 
reaction \cite{for94} due to limits in the energy range. A resonance at $5.13$~MeV would be 
a good candidate for the predicted $2^+$. 

The probability of populating a resonance in a transfer reaction is dependent on the overlap 
between the total wave function of the resonance and the configuration probed by the reaction. 
Take (t,p) as an example, assuming a direct reaction, the two-neutrons from a relative $s$-state 
in the triton would prefer an $s$-wave in the first Jacobi system. Comparing this configuration 
with the total wave function can be done by looking at the weights in tables~\ref{tab1}-\ref{tab3}.  
Here $0^{+}$ is dominated by $s$-wave of $60\%$, twice as large as $1^{-}$ with $30\%$ while 
$2^{+}$ is reduced by an additional factor of $3$ to about $12\%$. This indicates a larger probability 
of probing a $0^+$ than a $1^-$ and especially a $2^+$ resonance in a (t,p) reaction. These weights 
have been calculated for the bound states by Romero-Redondo et al \cite{rom07b} to be $90\%$, 
$37\%$, $46\%$ for he $0^+_1$, $2^+$, $1^-$ states. The weights predict a stronger population of 
the $0^+$ ground state and the bound $2^+$ than the population of the bound $1^-$. These findings 
are consistent with the (t,p) measurement of Fortune, \cite{for94}. The strong population of the 
two resonances in the experiment is then not favored by the suggested $2^+$ structure, which 
indicate only very weak population. The suggested quantum numbers of $0^+$ for the lowest resonance 
and $1^-$ for the second is more consistent with the large population of especially the lowest resonance. 

Sequential transfers through $^{11}$Be could also occur, and distort the picture. Whether the reaction 
is direct or sequential cannot be distinguished experimentally, but information from a (d,p) reaction 
might be helpful. A similar comparison can be made for a (d,p) reaction. Again assuming the simplest 
possible reaction, adding one neutron to the ground state of $^{11}$Be. In this case all weights are 
around $1\%$, which indicates that contributions from higher order reactions would be competitive. In 
fact for all three sets of quantum numbers $(0^+,1^-,2^+)$ an excitation to the $d$-shell is required 
to get more than $10\%$ weight, hence coupling to these continuum states should be taken into account 
in any reaction calculation. This is supported by a scattering experiment with $^{11}$Be on $^{64}$Zn 
which shows a large break up channel for $^{11}$Be \cite{dip10}. 

A (d,p) reaction could also be used to populate the predicted $0^-$ state around the one neutron threshold, 
as the reaction does not favor natural parity states. Comparing the strengths for the simplest reactions 
to populate a $0^-$ from a (t,p) and (d,p) reaction respectively can be done by comparing the weights in 
tab~\ref{tab4b}. Here the weights for the (t,p) is less than $1\%$ while it is $55\%$ for a (d,p). Therefore 
a population in a (t,p) should be very weak, nonetheless an indication of a broad weakly populated 
resonance at $-0.3$~MeV is seen in the (t,p) reaction by Fortune, \cite{for94}. The peak is to very 
weak, but if it is indeed a resonance, it should be populated and seen in a (d,p) measurement.

Further improved experimental investigations would help to test the validity of the model interpretation. This includes measurements of both population and decay of the channels.

\section{Summary and Conclusions}

We employ the technique of hyperspherical adiabatic expansion method
combined with complex coordinate scaling. We investigate the continuum
states, mostly resonances, of $^{12}$Be in a three-body model where
the constituents are two neutrons and $^{10}$Be.  The previously known
four bound states are fairly well described in such a model.  This
therefore already determines a successful set of two-body
interactions, which are now used in an extension to study low-lying
continuum states above the three-body breakup threshold.

The most probable angular momenta and parities are $0^{\pm}$, $1^{-}$
and $2^{+}$.  We first compute a series of the lowest adiabatic
potentials with these quantum numbers.  They exhibit lots of structure
and several potentials have attractive regions at small distance. This
already indicates possible bound states or resonances with
corresponding properties.  The energies are in principle found as
solutions to the hyperradial equations but the three-body model cannot
be expected to provide precise energies. Therefore we fine-tune with a
three-body potential which moves the bound states/resonances up or
down.  The widths of the resonances are then correlated with the
resonance position.  We should here remember that our computed width
should be larger than the measured value because we only have the
three-body component in the model. Any additional components would
tend to reduce the computed width.

The naturally occurring lowest of these states is the $0^{-}$ bound
state.  So far it has not been found experimentally and perhaps it
instead is pushed up above the one- or perhaps even the two-neutron
threshold. In attempts to move the energy from a bound to an unbound
state, we find that the width increases dramatically just above the
lowest one-neutron emission threshold.  The structure of the state
then corresponds to a coherent combination of $^{11}$Be in the ground
state surrounded by one neutron and a genuine three-body structure.
Quickly it becomes impossible to trace the resonance which could turn
into either a virtual state or a very broad neutron-$^{11}$Be resonance. 

Both $1^{-}$ and $0^{+}$ resonances can rather easily be placed at the
positions of the two lowest resonances at about $0.9$~MeV and $2.0$~MeV
above the three-body threshold.  It is then natural to associate these
quantum numbers with these resonances.  The width comparison to
measured values suggest that $0^{+}$ is the lowest state at about
$0.89$~MeV and $1^{-}$ is at $2.03$~MeV, but the opposite can not be
excluded.  The $2^{+}$ state also
appears as a resonance but it requires a very strong three-body
attraction to pull it down to these low energies.  It is then natural
to associate this state with a higher-lying resonance at about
$5.36$~MeV.

We calculate the structure of these resonances expressed as
partial-wave decomposed configurations. The neutron-neutron relative
wave functions are mixtures of $s$, $p$, and $d$-waves whereas the
neutron-$^{11}$Be relative wave functions consists of essentially only
$d$-waves for the $0^{+}$ and $2^{+}$ resonances, and both $p$ and
$d$-waves for the $1^{-}$ resonance.

The assignment of quantum numbers requires additional information. The
lowest-lying resonances have for some time tentatively been assigned
to $2^{+}$. The present model rather suggest $1^{-}$ and $0^{+}$. To
find further evidence we calculate the branching ratios for decays
into different channels. These quantities are observables and could
help to confirm assignments of angular momentum and parity.  We find
several coexisting decay channels for all three resonances depending
on which energy position is chosen.  Channels like one-neutron,
three-body direct as well as three-body sequential via different
$^{11}$Be resonances are all possible. 

The detailed comparison to measured results must unfortunately include
calculations of transfer cross sections. This would provide
information about population of the resonances which in combination
with our branching ratios could distinguish between different angular
momentum and parity assignments. 

In summary, we have calculated the low-lying three-body resonance
structure of $^{12}$Be. Partial wave decomposition for small as well
as large distances are now available for this model.  We find
surprisingly many different resonances and bound states for such a
light neutron dripline nucleus.  The computed structures can now be
compared to measurements. Our results and the available data are
all consistent with $0^+$ at $0.89$~MeV, $1^-$ at $2.03$~MeV and
$2^+$ at $5.13$~MeV.

\acknowledgments
This problem was suggested by H. Fynbo and
K. Riisager in connection with detailed analysis of experiments of
neutron transfer from deuteron to $^{11}$Be.  We appreciate the
information from numerous discussions.  This work was partly 
supported by funds provided by DGI of MINECO under contract 
No. FIS2008-01301.

\end{document}